\documentclass[twocolumn]{aastex631}

\newcommand{\mytilde}{\raise.19ex\hbox{$\scriptstyle\sim$}}
\usepackage{amsmath}
\shorttitle{SMACSJ0723 Weak Lensing}
\shortauthors{Finner et al.}

\begin{document}

\title{The First Weak-lensing Analysis with the James Webb Space Telescope: SMACS J0723.3–7327}

\correspondingauthor{Kyle Finner}
\email{kfinner@caltech.edu}

\author[0000-0002-4462-0709]{Kyle Finner}
\affiliation{IPAC, California Institute of Technology, 1200 E California Blvd., Pasadena, CA 91125, USA}

\author[0000-0002-4462-0709]{Andreas Faisst}
\affiliation{IPAC, California Institute of Technology, 1200 E California Blvd., Pasadena, CA 91125, USA}

\author[0000-0001-7583-0621]{Ranga-ram Chary}
\affiliation{IPAC, California Institute of Technology, 1200 E California Blvd., Pasadena, CA 91125, USA}

\author[0000-0002-5751-3697]{M. James Jee}
\affiliation{Department of Astronomy, Yonsei University, 50 Yonsei-ro, Seodaemun-gu, Seoul 03722, Republic of Korea}
\affiliation{Department of Physics and Astronomy, University of California, Davis, One Shields Avenue, Davis, CA 95616, USA}



\begin{abstract}
Utilizing the James Webb Space Telescope Early Release NIRCam Observations, we perform a weak-lensing analysis of the massive galaxy cluster SMACS J0723.3–7327 ($z=0.39$). We investigate the spatial variation of the PSF from the stars in the mosaic image. Our measurements show that the PSF for both modules has very small spatial and temporal variation with average complex ellipticity components of $e_1=0.007\pm0.001$ and $e_2=0.029\pm0.001$ in the observed north-up reference frame. We create PSF models through a principal component analysis of the stars and show that they properly account for the ellipticity of the PSF with residual shapes of $e_1=(0.3\pm3.5)\times10^{-4}$ and $e_2=(1.8\pm4.0)\times10^{-4}$. We select background galaxies by their photometric redshift and measure galaxy shapes by model fitting. Our weak-lensing source catalog achieves \mytilde 215 galaxies arcmin$^{-2}$. We map the convergence field of SMACSJ0723 and detect the cluster with a peak significance of $12.2\sigma$. The mass distribution is found to elongate in the east-west direction with an extension to the northeast edge of the field of view where a candidate substructure is found in the \textit{Chandra} X-ray imaging. We fit the tangential shear with a Navarro-Frenk-White model and estimate the mass of the cluster to be $M_{500}=7.9\pm1.1\times10^{14}$ M$_{\odot}$ ($M_{200}=11.4\pm1.5\times10^{14}$ M$_\odot$ ), which agrees with existing mass estimates. Combining the multiwavelength evidence from literature with our weak-lensing analysis, we hypothesize that SMACSJ0723 is observed near first pericenter passage and we identify candidate radio relics.

\end{abstract}



\section{Introduction} \label{sec:intro}
Weak gravitational lensing is an ideal tool for measuring the matter distribution of the universe. It has been applied frequently to study galaxy clusters and can provide mass estimates that do not require an assumption of hydrostatic equilibrium nor a scaling from gas to total mass. Thus, it is particularly useful for studying the mass of unrelaxed systems.

SMACS J0723.3–7327 (hereafter SMACSJ0723) is a galaxy cluster at a redshift of 0.39 with equatorial coordinates 07h23m13s, -73d27m25s. It was discovered in the Massive Cluster Survey \citep{2001ebeling} from its prominent X-ray emission. The mass of the cluster was reported in the Planck Sunyaev-Zel'dovich Effect (SZE) study to be $M_{500}=8.39\pm0.39\times10^{14}$ M$_\odot$ \citep{2016planck} with an SZE signal-to-noise (S/N) of 15.7. \cite{2016martinez, 2018martinez} reported the presence of a giant radio halo detected in the $1.1-3.1$~GHz channels of the Australia Telescope Compact Array (ATCA). Giant radio halos \citep{1977jaffe, 2006cassano} are diffuse synchrotron emission that is thought to be from the re-acceleration of relativistic electrons arising from merger-induced turbulence in the intracluster medium (ICM). Thus, SMACSJ0723 has signs of being an unrelaxed system. \cite{2020lovisari} included SMACSJ0723 in their study of archival XMM-\textit{Newton} observations and estimated the mass to be $M_{500}=10.10^{+1.57}_{-1.23}\times10^{14}$ M$_\odot$ by fitting the density and temperature profiles of the ICM. 

The launch of the JWST and the subsequent release of the Early Release Observations piqued interest in SMACSJ0723. Five strong-lensing studies \citep{2022mahler, 2022caminha, 2022pascale, 2022golubchik, 2023Sharon} of the cluster have since been published with all of them coming to a consensus that the lensing critical curves elongate in the east-west direction with an effective Einstein radius of approximately $15\arcsec$ for a source at $z_s=1.45$ \citep{2022golubchik} to $25\arcsec$ for a source at $z\rightarrow\infty$ \citep{2022caminha}. \cite{2022liu} combined the power of eROSITA and \textit{Chandra} to investigate X-ray emission from SMACSJ0723. They showed that the ICM is hot (\mytilde 10~keV) and estimated the mass to be $M_{500}=9.79\pm5.07\times10^{14}$ M$_\odot$. In addition, their study of the X-ray emission notes that the X-ray brightness peak is cospatial with the brightest cluster galaxy (BCG) but has an elongation of the core to the west of the BCG, which may be from sloshing of the ICM \citep[for more on sloshing see][]{2007markevitch, 2011zuhone}. They also report a gas clump to the northeast but suggest that it is too faint to have significantly altered the hydrostatic state of the cluster.

With the elevated interest in SMACSJ0723 as a gravitational lens, we aim to measure the weak-lensing signal and estimate the total mass of the cluster. Weak lensing is an ideal tool for measuring the total mass because it is not confined to the inner region of the cluster nor is it subject to a hydrostatic equilibrium assumption. A summary of the JWST observations and data reduction are provided in Section \ref{sec:observations}. We provide an overview of the weak-lensing theory and our technique in Section \ref{sec:wl}. Since no previous weak-lensing analysis with the JWST exists, we characterize the point-spread function (PSF) and create models (Section \ref{sec:psf}). In Section \ref{sec:source_selection}, we carefully select the background galaxies. The mass maps and mass estimates of our study are presented in Section \ref{sec:results}. We discuss the results and conclusions in Sections \ref{sec:discussion} and \ref{sec:conclusions}.

For this work, we assume a flat $\Lambda$CDM cosmology with $H_0=70$ km Mpc$^{-1}$ s$^{-1}$, $\Omega_m=0.3$, and $\Omega_\Lambda=0.7$. At the redshift of SMACSJ0732 $(z=0.39)$, $1\arcsec=5.34$ kpc. All magnitudes are reported in the AB magnitude system. Masses are reported as $M_{500(200)}$, where the average density within a spherical halo of radius $R_{500(200)}$ is 500 (200) times the critical density of the universe.  

\section{JWST Imaging}\label{sec:observations}
SMACSJ0723 was observed with the JWST NIRCam on 07 June 2023 (PI Pontoppidan). The observations were carried out in the F090W, F150W, and F200W shortwave (SW) filters and the F277W, F356W, and F444W longwave (LW) filters with integration time of 7537s in each filter. NIRCam has two modules called A and B. For the imaging of SMACSJ0723, module B was centered on the BCG with the V3 position angle rotated 145 degrees east of north, which positioned module A to the southwest. The imaging was reduced with the Grizli pipeline \citep{2019grizli} and coadded into mosaics for each filter with a pixel scale of $0\farcs02$.

For our weak-lensing analysis, we selected the F200W image because it has the best sampling rate (PSF size to pixel scale ratio). \cite{2023finnerb} and \cite{2021kannawadi} demonstrate the importance of having a sampling rate that is near the Nyquist criteria \citep{Shannon1949} when measuring galaxy shapes. In addition, since F200W is the reddest SW filter, it is more sensitive to the distant galaxies than the bluer SW filters, which benefits weak lensing because the more distant galaxies are more affected by lensing (as shown in Equation \ref{eq:kappa}). Furthermore, the galaxies seen through redder filters have smoother light profiles because they are less affected by dust extinction in the sources and trace the older stellar population \citep{2018lee}. 

We ran SExtractor \citep{1996bertin} in dual-image mode with F200W as the detection image and created photometric catalogs for the F090W and F200W filters. The SExtractor settings that we used are \texttt{DETECT\_MINAREA}=5, \texttt{DETECT\_THRESH}=2, \texttt{DEBLEND\_NTHRESH}=16, and \texttt{DEBLEND\_MINCONT}=0.01.

\section{weak-lensing and Shape Measurement Theory}\label{sec:wl}
The weak-lensing effect imparts small distortions to the background galaxy images. The distortions are the anistropic shear $\gamma$ and the isotropic convergence $\kappa$. The shear transforms the ellipticity of galaxy images while the convergence magnifies galaxy images. The transformation is described by the Jacobi matrix:

\begin{equation}
    A = \left(1-\kappa\right)\left(\begin{matrix} 1-g_1 & -g_2 \\ -g_2 & 1+g_1 \end{matrix}\right),
\end{equation}
where $g_i=\gamma_i/(1-\kappa)$ is the complex reduced shear. The convergence $\kappa$ is the projected mass density $\Sigma$ normalized by the critical surface density $\Sigma_c$:

\begin{equation} \label{eq:kappa}
    \kappa = \frac{\Sigma}{\Sigma_c} ; \Sigma_c = \frac{c^2 D_s}{4\pi G D_{ls} D_l},
\end{equation}
where $c$ is the speed of light, $G$ is the gravitational constant, and $D_{ls}$ (lens to source), $D_l$ (lens), and $D_s$ (source) are angular diameter distances. The ratio $\beta = D_{ls}$/$D_s$ is referred to as the lensing efficiency. 

We measure the ellipticity of galaxies by fitting a PSF-convolved Gaussian distribution to the surface brightness. Of the seven parameters of the elliptical Gaussian equation, we fix the centroid ($x,y$) and the background to the SExtractor values. The remaining four parameters that are fit are: amplitude, semi-major axis $a$, semi-minor axis $b$, and position angle $\theta$. From the measurement, the complex ellipticity $e=e_1+ie_2$ is calculated

\begin{align}
\centering
e_1 &= \frac{a-b}{a+b} \cos 2 \theta, \\
e_2 &= \frac{a-b}{a+b} \sin 2 \theta.  
\end{align}

As with any model fitting method for measuring the shapes of galaxies, elliptical Gaussian fitting is subject to model bias \citep{2018mandelbaum}. We use simulations to understand the multiplicative and additive biases that affect our weak-lensing pipeline. This shear calibration technique, called SFIT, was shown to be the most effective in the GREAT3 challenge \citep{2015mandelbaum}. Through our image simulations, we find multiplicative bias factors of 1.11 and 1.07 for $e_1$ and $e_2$, respectively. For the additive bias, our simulations find that they are negligible ($<10^{-3}$) for this study of the massive cluster SMACSJ0723.

\section{Point-Spread Function Investigation and Modeling} \label{sec:psf}

\begin{figure}[tb]
    \centering
    \includegraphics[width=0.48\textwidth]{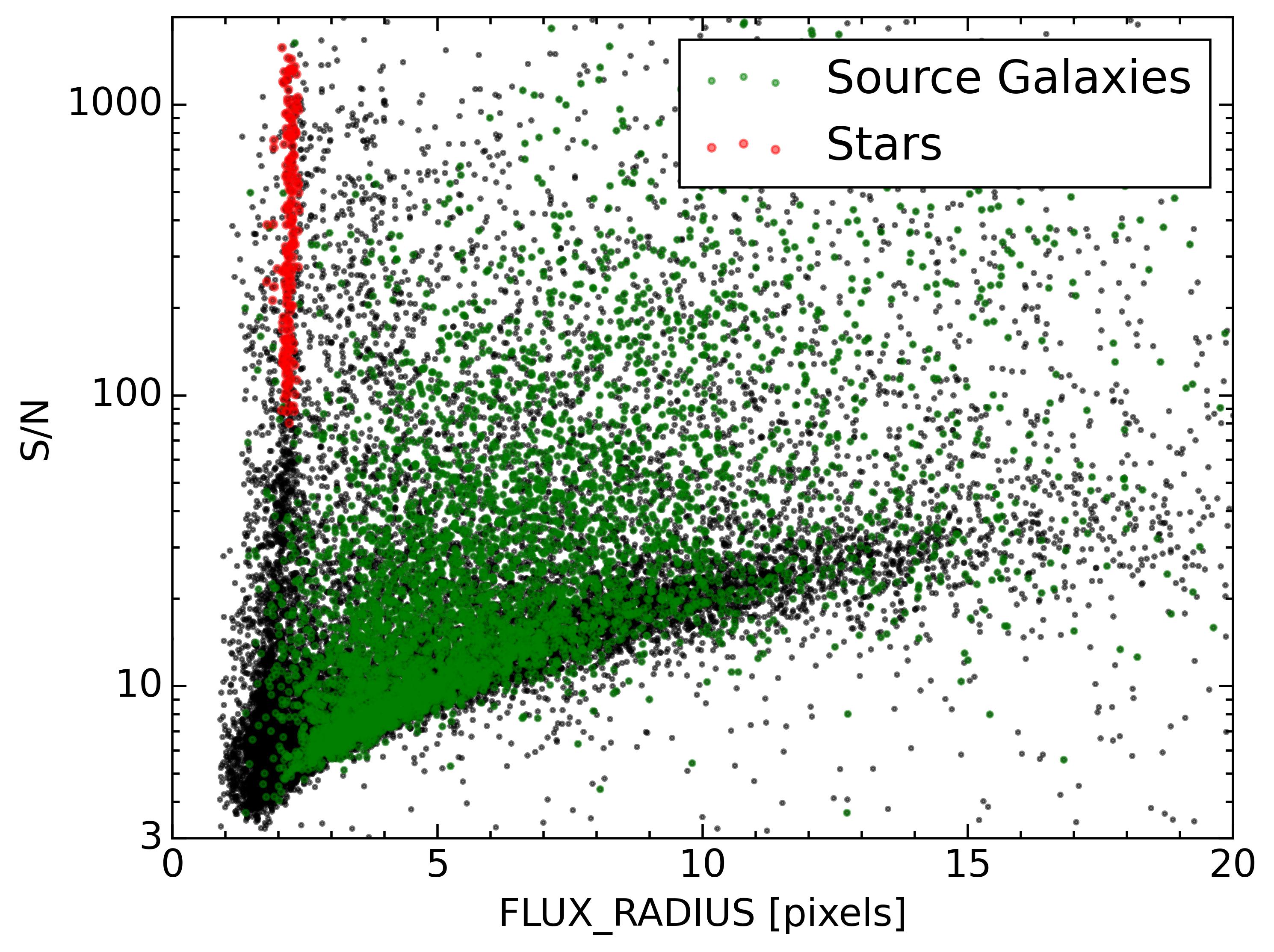}
    \caption{S/N as a function of SExtractor \texttt{FLUX\_RADIUS} for all objects (black circles) in the SMACSJ0723 mosaic. The red circles are the stars selected for PSF modeling, which follow a stellar locus that manifests because stars are effectively point sources and take on the size of the PSF. The green circles are the source galaxies that are used for weak-lensing analysis (Section \ref{sec:source_selection}).}
    \label{fig:psf_stars}
\end{figure}

\begin{figure*}[tb]
    \centering
    \includegraphics[width=\textwidth]{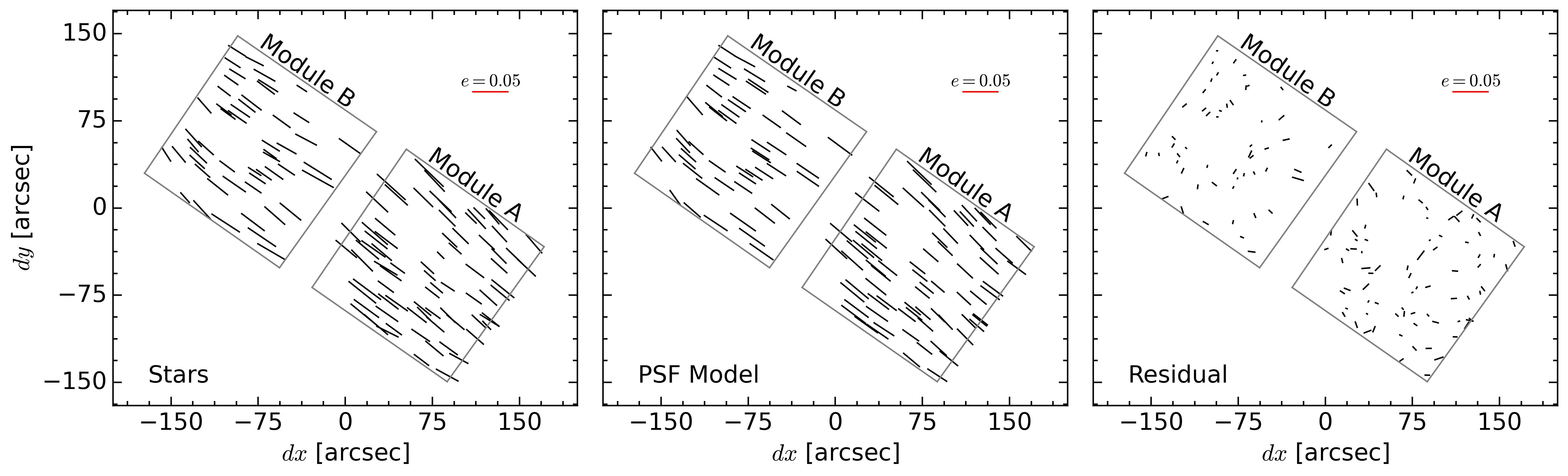}
    \caption{The whiskers represent the ellipticity direction and magnitude from quadrupole measurements. Left: Stars as selected in Section \ref{sec:psf} and displayed in Figure \ref{fig:psf_stars}. Middle: PSF model derived from a PCA of the stars. Right: Residual from subtracting the PSF model ellipticities from the star ellipticities (i.e., left panel minus middle panel). An ellipticity of 0.05 is displayed in the top right of each panel and all panels are on the same scale.}
    \label{fig:star_ellip}
\end{figure*}

\begin{figure}[ht]
    \centering
    \includegraphics[width=0.45\textwidth]{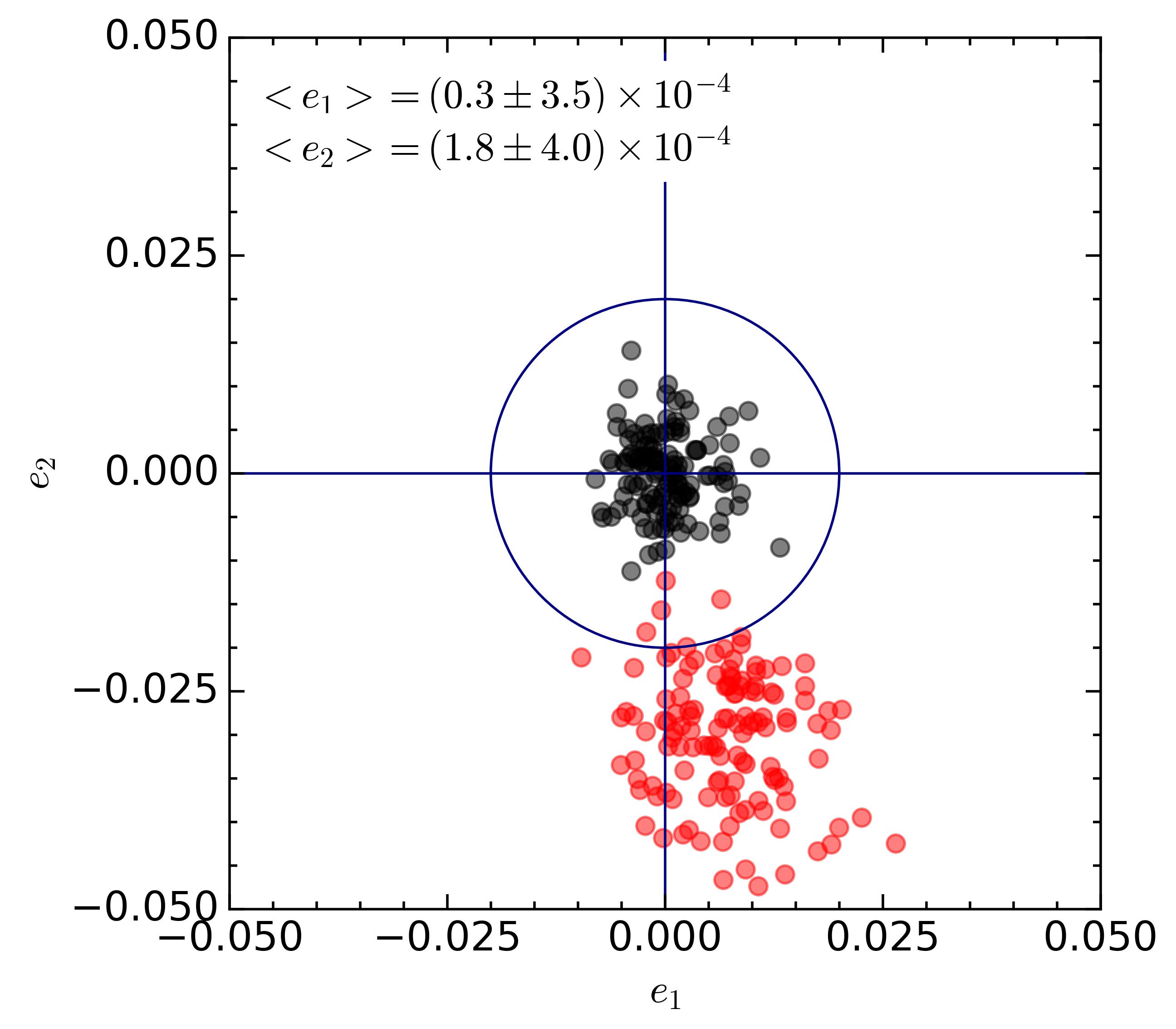}\\
     \includegraphics[width=0.45\textwidth]{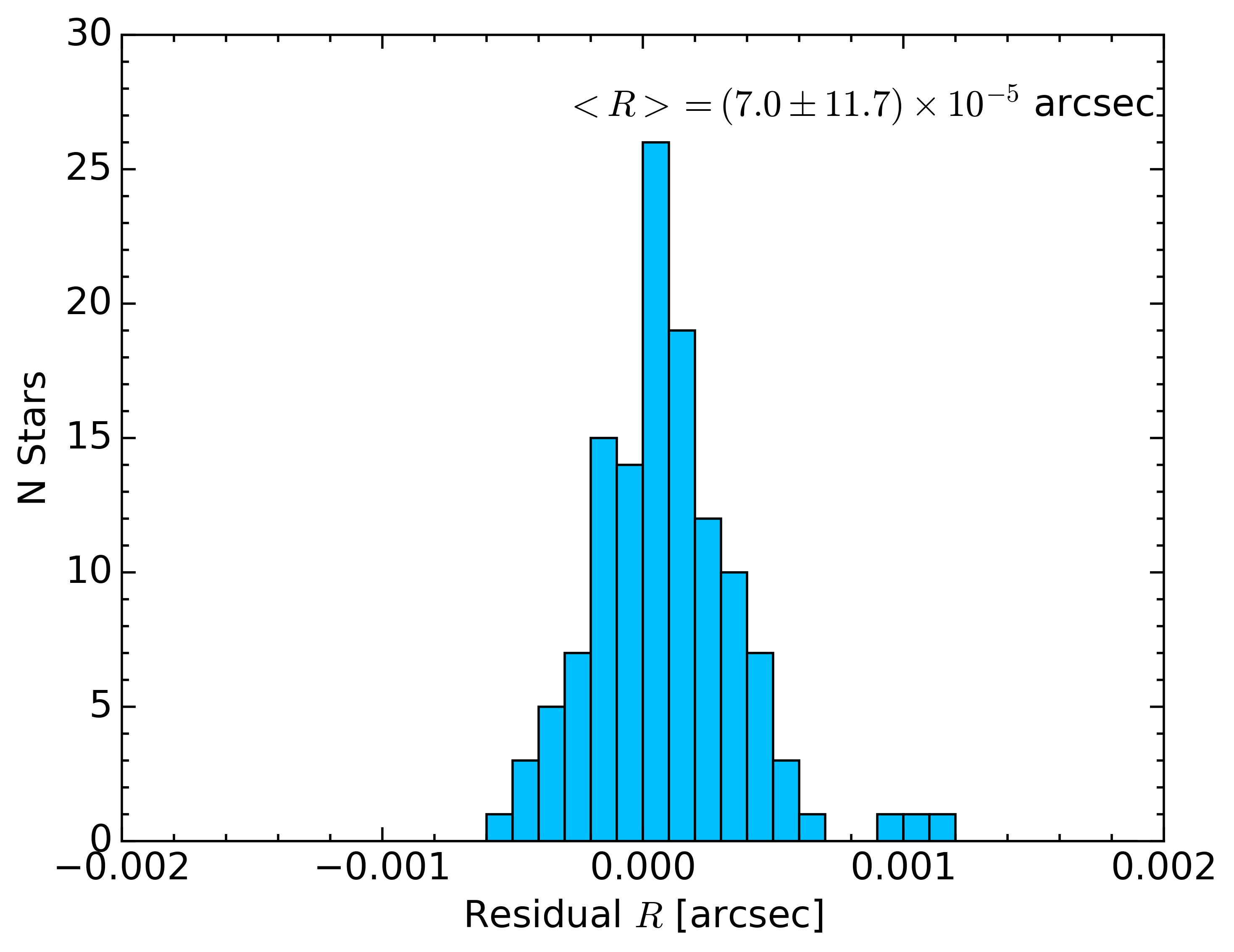}
    
    \caption{PSF model corrections. Top: Complex ellipticity components of the stars (red circles) and residual (black circles). The PSF models correct for the ellipticity and the spatial variation (scatter) of the PSF. The median and standard error for the residuals are reported in the top left. Bottom: Residual size $R$ of the PSF after subtracting the PSF model size from the size of stars. }
    \label{fig:psf_residuals}
\end{figure}
The PSF must be modeled and its effect removed because the distortion caused by the PSF mimics the weak-lensing distortion. We investigate the spatial dependence of the PSF by selecting stars from the mosaic based on their size and magnitude properties (Figure \ref{fig:psf_stars}). The stellar locus is well defined at a \texttt{FLUX\_RADIUS} of \mytilde 2 pixels (SExtractor-measured FWHM\mytilde $0\farcs07$). We select stars (red circles) by grid searching the size-S/N relation for the densest region in a S/N range of 80 $<$ S/N $<$ 2000. After finding the densest region, we perform a linear fit and select objects within $\pm0.2$ of the linear fit. The S/N constraint is used to exclude stars that are too faint to be reliable or too bright and saturated. In addition, we require the measured centroid of each star to be within 0.5 pixels of its SExtractor position (\texttt{XWIN\_IMAGE}, \texttt{YWIN\_IMAGE}), which filters out stars with bright neighbors. This selects 129 PSF stars in total with about 65 per module. 

One of the systematics that should be considered when modeling the PSF is the brighter-fatter effect \citep{2014antilogus, 2018plazas}. We find that the FWHM of the stars that we have selected increases by $\mytilde2\%$ from 26th magnitude ($S/N\mytilde80$) to 20th magnitude ($S/N\mytilde2000$). We modeled the impact of the brighter-fatter effect on weak-lensing measurements in \cite{2020finner}. From our model, we expect a $2\%$ overestimated PSF size to induce a multiplicative bias to galaxy shape measurements of $\lesssim1\%$. This level of multiplicative bias will not have a significant impact on our mass estimate of SMACSJ0723. However, our investigation of the brighter-fatter effect only considers the 129 stars that we are able to detect in the cluster field. A more thorough analysis should be completed on globular cluster imaging.      

To investigate the ellipticity of the PSF, the PSF stars are measured with the quadrupole moment equation \citep[for more on quadrupole moments see][]{2014mandelbaum}, where we use a circular Gaussian weighting with a standard deviation of 3 pixels to suppress noise during the measurement. The left panel of Figure \ref{fig:star_ellip} presents the ellipticity of the stars. The ellipticity of the stars, or equivalently the PSF, is elongated primarily in the horizontal direction of the detector axes for both modules. The quadrupole measurements find that the mean PSF ellipticity components are $e_1=0.007\pm0.001$ and $e_2=0.029\pm0.001$. In the drizzled frame, where north is up and west is right, the ellipticity is dominated by the $e_2$ term because of the V3 position angle of the imaging. Furthermore, we measured the ellipticities of the same stars in each component frame that comprises the coadded mosaic and found that scatter of the mean for $e_1$ and $e_2$ is \mytilde0.001 within the 8 visits of the telescope. Thus, the PSF does not vary significantly throughout the two hour observation.

For the 129 PSF stars, we record their positions ($x,y$), derive a mean star, and stack the stars into a 129-stars by 961-pixels (31 by 31 px flattened) array. The mean star is subtracted from each of the stars in the array to derive the residual array. We perform a principal component (PC) analysis of the residual array. By analyzing the PCs sorted by decreasing variance, we decide to keep the first 21 PCs, which contain \mytilde90\% of the variance. Our tests find that including fewer PCs causes an increase in scatter in the residual ellipticities and including more PCs leads to a poorer ellipticity correction. The result of the PC analysis is the residual array transformed to the 21 PC basis.

To retrieve the PSF model at a given $x,y$ coordinate, we fit the transformed residual array with a sixth order polynomial and use the coefficients to reconstruct a 2D image of the residual at the $x,y$ coordinates. That 2D image is added to the mean star to obtain the PSF model. Variations of this PSF modeling recipe have been validated and applied to HST observations with ACS \citep{2007bjee}, WFC3-IR \citep{2020finner} and ground-based observations \citep{2013jee, 2017finner, 2020hyeonghan}.

As a test of the PSF modeling capabilities, we determined the PSF model for each of the stars. The middle panel of Figure \ref{fig:star_ellip} presents the quadrupole measured ellipticities of the PSF models that are designed for the stars from the left panel. It is apparent that the PSF models are mimicking the ellipticity of the stars. The right panel is the subtraction of the PSF model ellipticity from the star ellipticity. It shows that the ellipticity of the PSF has been corrected and small, randomly oriented residuals remain. 

The top panel of Figure \ref{fig:psf_residuals} quantifies the ellipticity of the stars (red circles) that were presented in the left panel of Figure \ref{fig:star_ellip}. The ellipticity of the stars is corrected and the residual ellipticity (black circles) is centered near 0,0 with a reduced scatter. The mean residual ellipticities are $<$$e_1$$>\ =(0.3\pm3.5)\times10^{-4}$ and $<$$e_2$$>\ =(1.8\pm4.0)\times10^{-4}$. The bottom panel shows the residual size $R$ after subtracting the PSF model sizes from the star sizes. The residual sizes have a median and standard error of $<$$R$$>\ =(7.0\pm11.7)\times10^{-2}$ mas. The residual ellipticity and size will cause an additive and multiplicative bias in our shape measurements, respectively. However, as we have shown, the magnitude of these residuals are small and will have a negligible impact on our study of a massive galaxy cluster.

\section{Source Selection}\label{sec:source_selection}
Source selection is a critical step in weak lensing because only the background galaxies carry the weak-lensing signal. In many studies of galaxy clusters, the background galaxies are selected from their color and magnitude \citep[for example][]{2022zohren}. However, SMACSJ0723 has extensive multi-band photometry that enables the measurement of photometric redshifts (photo-zs). The existence of photo-zs simplifies the selection of background galaxies for our weak-lensing analysis. 

We create a source catalog by cross matching our shape catalog with the existing photo-z catalog \citep{gabriel_brammer_2022_6874301} and enforcing the following constraints. To safely select galaxies that are behind the cluster, we constrain the catalog to galaxies with a $1\sigma$ lower limit on their photo-z probability distribution to greater than 0.45. This constraint extends behind the cluster by about 6 times the typical line-of-sight velocity dispersion ($\delta z\mytilde0.01$) of a massive merging cluster \citep{2019golovich} and limits contamination by foreground while preserving a high number density of background galaxies. We also constrain the F200W magnitude to fainter than 24, which aids in removing foreground and cluster galaxies. To select galaxies with reliable photo-z estimates, only those with photo-z uncertainty less than 25\% ($\delta z\mytilde 0.3$ at $z=1.3$) are kept. Source galaxies are also selected based on the properties of their shape measurement. We limit the catalog to galaxies with ellipticity uncertainty $\delta e<0.3$ and reject objects that have an ellipticity that is greater than 0.85 because they tend to be spurious detections, such as diffraction spikes from stars. Finally, we remove other spurious objects by eye, such as objects that overlap with diffraction spikes, because they have compromised shapes. After applying the constraints, the remaining source catalog contains 2108 galaxies across the approximately 9.7 arcmin$^{2}$ field of view, which gives a source density of \mytilde 215 galaxies arcmin$^{-2}$. This source density is unprecedentedly high in comparison to typical HST ACS analyses that achieve 50 galaxies arcmin$^{-2}$ or ground-based analyses that reach 30 galaxies arcmin$^{-2}$.

\begin{figure}
    \centering
    \includegraphics[width=0.48\textwidth]{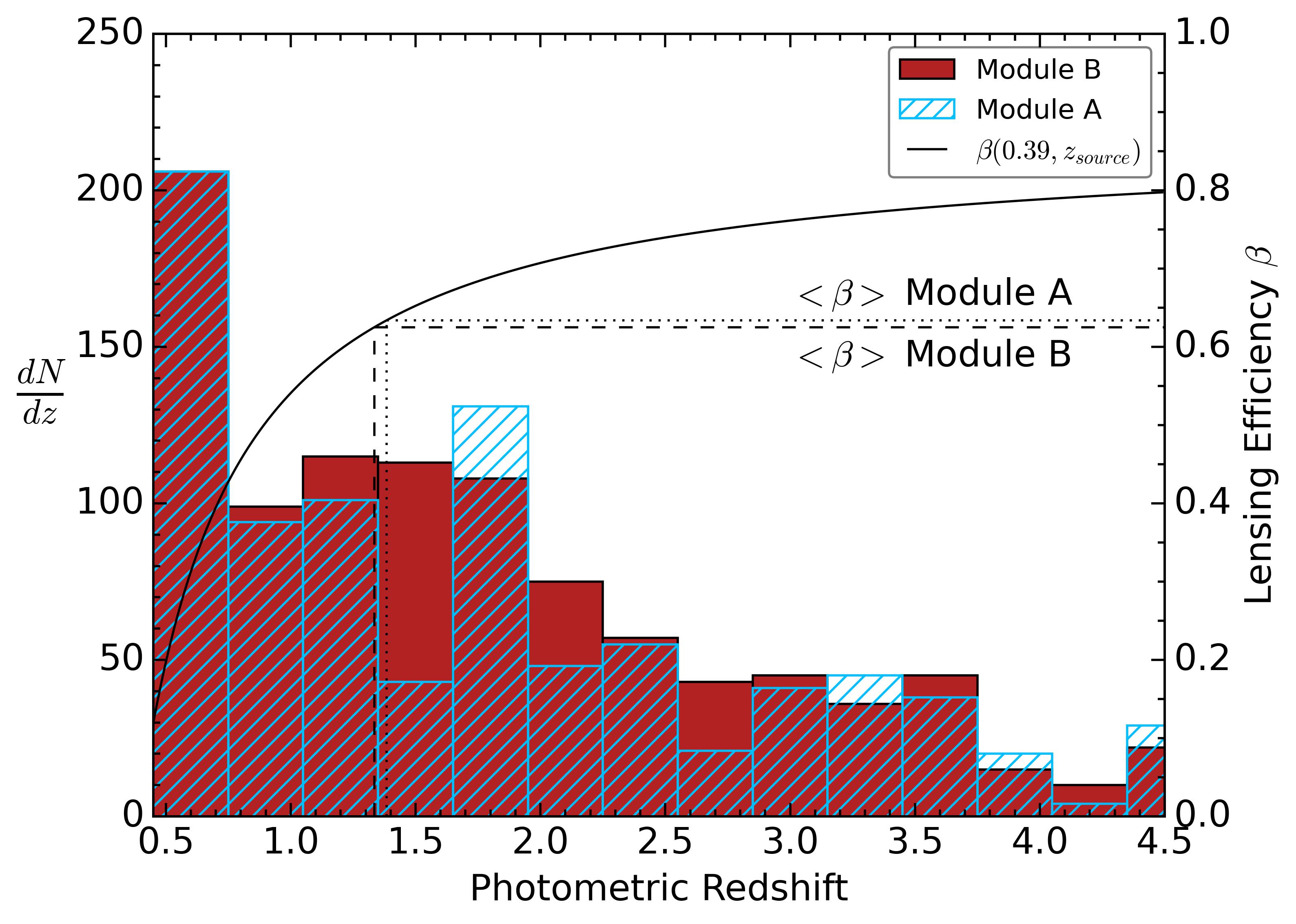}
    \caption{Left axis: photometric redshift distribution for the source catalog separated by module. Right axis: lensing efficiency $\beta$ as a function of source redshift for SMACSJ0723 ($z=0.39$). The horizontal dashed and dotted lines mark the average $\beta$ for each module and the vertical versions mark the mean redshift. Histograms, $\beta$ values, and mean redshifts are similar for the two modules.}
    \label{fig:dndz}
\end{figure}

The photo-z catalog that we have utilized in source selection was created with the Grizli pipeline using the existing HST ACS (F435W, F606W, and F814W) and JWST imaging. However, the HST imaging is confined within the ACS window ($3.4\arcmin$ by $3.4\arcmin$) that is centered on the BCG, while the JWST has two $2.2\arcmin$ by $2.2\arcmin$ modules. Thus, ACS observations completely cover the JWST module B but only partially cover module A. Figure \ref{fig:dndz} presents the redshift distribution of source galaxies separated by module. We find that the redshift distributions for the two modules are similar, which is indicative that the difference in filters is not causing a bias between the modules. Furthermore, we calculate the mean $\beta$ values (Equation \ref{eq:kappa}) for each module and find them to be similar with $\left<\beta_A\right>=0.634$ (horizontal dotted line) and $\left<\beta_B\right>=0.625$ (horizontal dashed line).    

\begin{figure}
    \centering
    \includegraphics[width=0.48\textwidth]{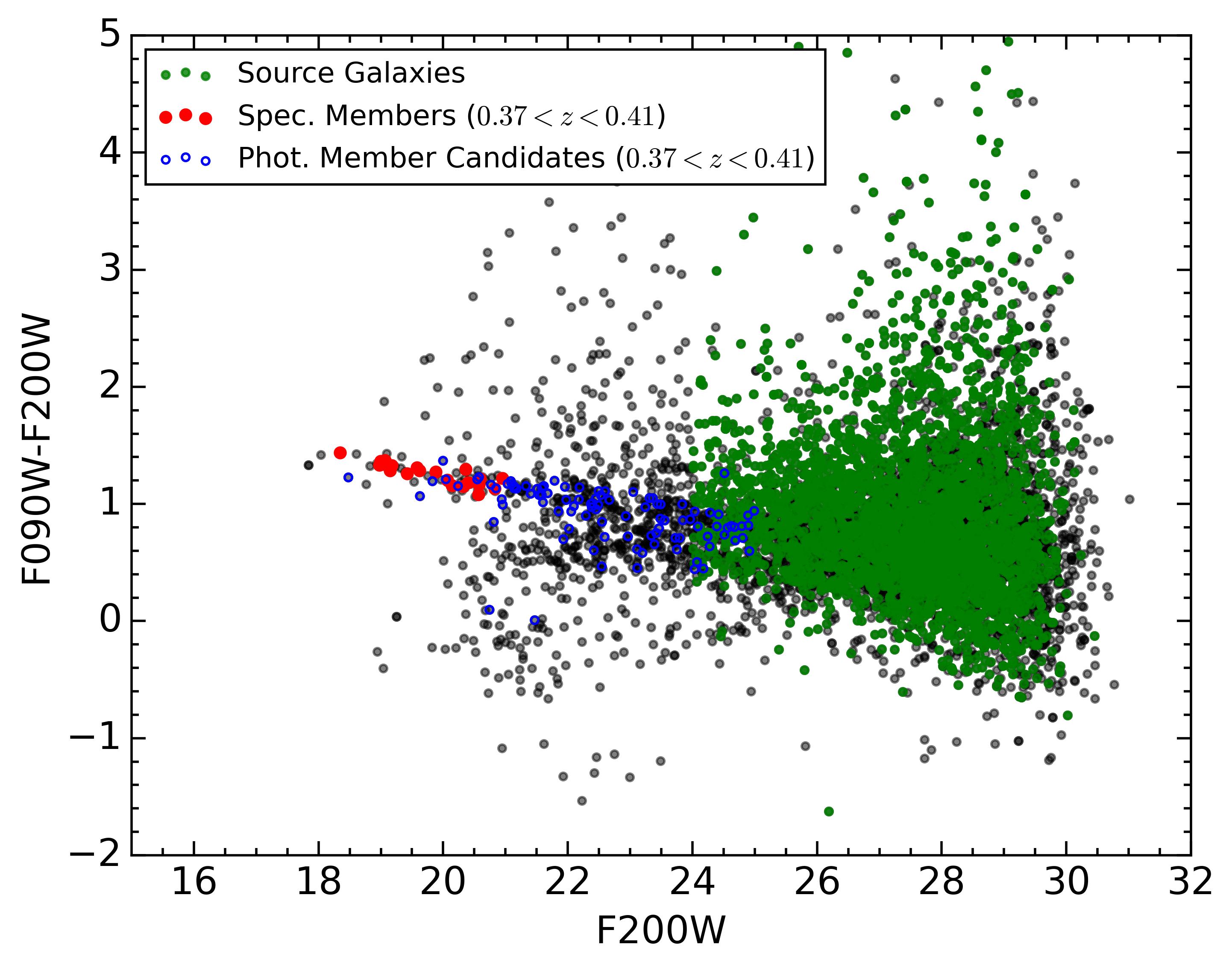}
    \caption{Color-magnitude diagram for SMACSJ0723. The filled red circles mark the spectroscopically confirmed cluster galaxies, which show a well-defined red sequence. The empty blue circles mark the photometrically selected cluster galaxies. The green circles are the source galaxies as selected in Section \ref{sec:source_selection}. The primary constraint on distance for source selection is the photo-z. The source catalog contains $215$ galaxies arcmin$^{-2}$. }
    \label{fig:cmd}
\end{figure}

Figure \ref{fig:cmd} presents the color-magnitude diagram (F200W magnitude vs. F090W-F200W color) for SMACSJ0723. Galaxies with spectroscopic redshift in the range $0.37<z<0.41$ (filled red circles) trace the tight red sequence of the galaxy cluster, which is likely prominent for the cluster galaxies because of the presence of the $1.6~\mu$m bump \citep{2002sawicki} in the F200W filter. Selecting galaxies with photo-z in the same redshift range, brighter than 25th magnitude, and photo-z uncertainty less than $25\%$ (open blue circles) extends the red sequence to fainter magnitudes but with increased scatter. The spec- and photo-z selected cluster galaxies are saved for mapping the stellar light of the galaxy cluster in Section \ref{sec:discussion}. The green circles in the figure are the source galaxies that are used for our weak-lensing analysis.   

\section{Results} \label{sec:results}
\subsection{Projected Mass Distribution}
\begin{figure*}
    \centering
    \includegraphics[width=\textwidth]{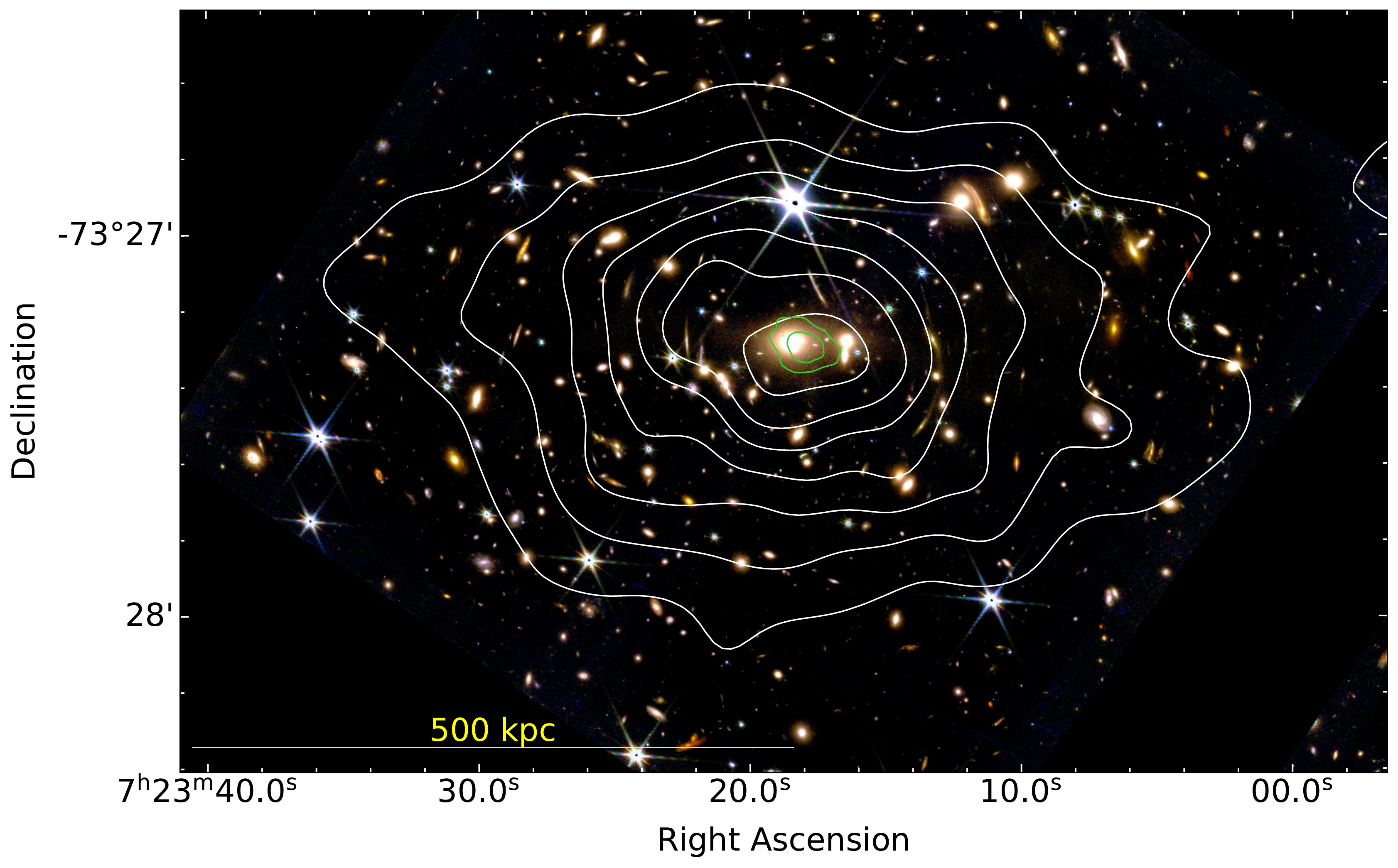}
    \caption{SMACSJ0723 convergence map (contours) overlayed on the JWST color-composite image (F105W, F200W, F277W). Contours start at the $1\sigma$ level and increase in $2\sigma$ steps. The peak of convergence is detected at the $12.2\sigma$ level and is spatially consistent with the BCG at the $1\sigma$ level (green contour). The convergence is elongated in the east-west direction with a slight extension in the northeast. }
    \label{fig:mass_map}
\end{figure*}
A primary goal of a weak-lensing analysis of a galaxy cluster is to map the dark matter distribution. We use a maximum entropy (MAXENT) technique \citep{1998seitz, 2000bridle, 2002marshall, 2007ajee} to reconstruct the projected mass distribution (mass map). The inputs to MAXENT are the positions, complex ellipticity components, ellipticity uncertainties, and redshifts of the galaxies in the source catalog. We generate 1000 realizations of the mass map from bootstraps of the source catalog and divide the original mass map by the standard deviation of the 1000 realizations to create a S/N map.

The white contours in Figure \ref{fig:mass_map} present the projected mass distribution. The contours start at $1\sigma$ and increase at $2\sigma$ intervals to a peak significance of $12.2\sigma$. The peak of the mass map is cospatial with the BCG. The morphology of the mass map is elongated in the east-west direction. In addition, a low-significance extension to the northeast is found. 

To test the significance of the mass peak agreement with the BCG, we fit the centroid of each mass map realization and calculate the $1\sigma$ and $2\sigma$ uncertainties of the distribution of mass peaks. The $1\sigma$ uncertainty has a diameter of approximately $6\arcsec$ thanks to the high galaxy source density that JWST imaging provides. These distributions (green contours on Figure \ref{fig:mass_map}) show an agreement between the mass peak and the BCG at the $1\sigma$ level.  

\subsection{Mass Estimation}
The mass of a galaxy cluster can be estimated by fitting a model to the tangential shear signal.  The tangential and cross shear are:

\begin{align}
\centering
    g_t &= -g_1\cos 2\phi - g_2 \sin 2\phi, \\
    g_x &= \ g_1 \sin 2\phi - g_2 \cos 2\phi,
\end{align}
where $g$ are the components of the reduced shear and $\phi$ is the angle measured from the horizontal axis that intersects the cluster center. Since the peak of the mass distribution is cospatial with the BCG and the X-ray brightness peak, we use the BCG as the center because it is the most precisely defined. 

\begin{figure}
    \centering
    \includegraphics[width=0.48\textwidth]{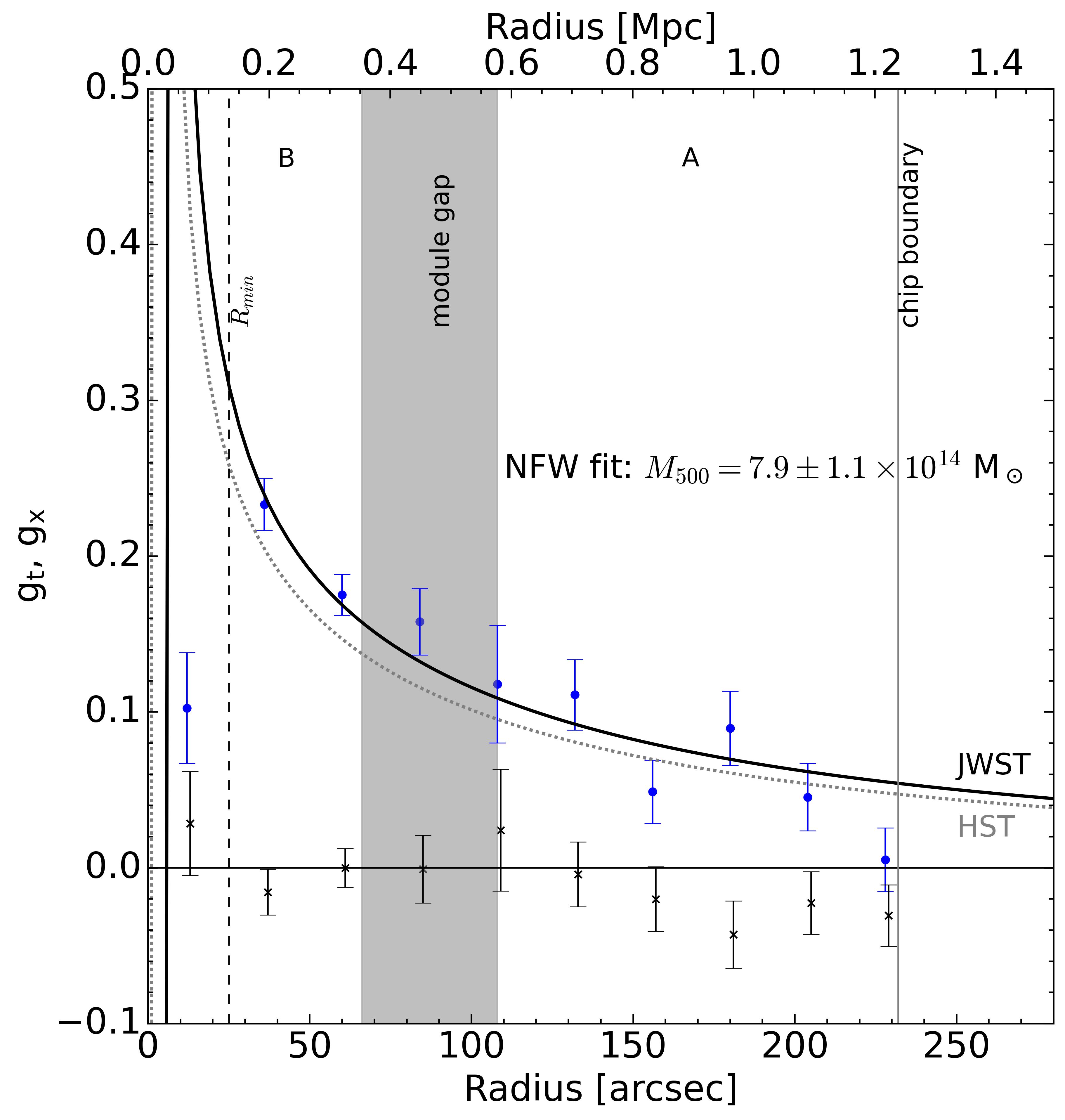}
    \caption{Radial dependence of the tangential shear (blue circles) and the cross shear (black $\times$'s). The errorbars are the standard errors, which include the Poisson noise. The vertical black dashed line is the minimum radius ($R_{min}$) of the fit and the maximum radius is shown at the chip boundary (vertical gray line). An NFW fit to the unbinned tangential shear is presented as the black curve. We have also overlaid the NFW fit from our HST analysis (gray dotted curve) from Appendix \ref{app:A}. }
    \label{fig:tan_shear}
\end{figure}

Figure \ref{fig:tan_shear} presents the tangential shear (blue circles), which is calculated by averaging the galaxy shapes in radial bins. The two modules of the JWST NIRCam are beneficial for measuring the tangential shear because the module that is not centered on the cluster, module A in this case, probes the outer regions of the galaxy cluster. However, when using the two modules an assumption of azimuthal symmetry is required. As expected for a massive galaxy cluster, the tangential shear near the cluster core is prominent ($g_t\sim 0.25$) and it remains prominent at radii $>110\arcsec$, where module A is situated. However, the statistical significance of the tangential shear at these large radii demonstrates how remarkable the JWST is for probing the dark matter distributions of galaxy clusters. 

We fit a shear model \citep{2000wright} that follows a Navarro-Frenk-White \citep[NFW;][]{1997navarro} mass profile to the unbinned tangential shear. We fix the mass-concentration relation to that of \cite{2019diemer} because the field of view does not extend to $R_{500}$. Furthermore, we omit the region inside the estimated effective Einstein radius (dashed line $R_{min}=25\arcsec$ in Figure \ref{fig:tan_shear}) of the tangential shear to bypass the strong-lensing regime where the weak-lensing assumption is violated. Our NFW fit estimates the mass of the cluster to be $M_{500}=7.9\pm1.1\times10^{14}$ M$_{\odot}$ where the statistical uncertainty is $\pm0.5$ and the systematic uncertainty propagated from the photo-z uncertainty contributes $\pm0.6$. We also calculate the mass at 200 times the critical density of the universe to be $M_{200}=11.4\pm1.5\times10^{14}$ M$_\odot$. 

\section{Discussion} \label{sec:discussion}
\subsection{Mass Consistency}
The mass of SMACSJ0723 (a.k.a. PSZ2 G284.97-23.69) was measured via the Sunyaev-Zel'dovich effect and found to be $M_{500}=8.39\pm0.39\times10^{14}$ M$_\odot$ \citep{2016planck}. \cite{2020lovisari} measured $M_{500}$ from the XMM-Newton observations of SMACSJ0723 by assuming the cluster is in hydrostatic equilibrium. Their mass estimate is $M_{500}=10.10^{+1.57}_{-1.23}\times10^{14}$ M$_\odot$. Equipped with eROSITA and \textit{Chandra} observations of SMACSJ0723, \cite{2022liu} estimated $M_{500}=9.79\pm5.07\times10^{14}$ M$_\odot$ and $M_{500}=9.15\pm4.65\times10^{14}$ M$_\odot$ for eROSITA and \textit{Chandra}, respectively. There are also mass estimates from strong-lensing analysis that measure the mass within the Einstein radius. We forego making a comparison to these mass estimates to prevent extrapolating them to a total mass for the cluster.

By fitting an NFW profile to the tangential shear, we measured the mass of SMACSJ0723 to be $M_{500}=7.9\pm1.1\times10^{14}$ M$_\odot$. This mass estimate is within $1\sigma$ of all mass estimates from the ICM emission. 

In Appendix \ref{app:A}, we perform a weak-lensing analysis of the archival HST ACS F814W imaging of SMACSJ0723. Following the procedure to fit the NFW model to the tangential shear from Section \ref{sec:results}, we determine a mass estimate of $M_{500}=6.5\pm1.7\times10^{14}$ M$_\odot$ from the HST. This mass estimate is consistent with the JWST estimate at the $1\sigma$ level but has a larger uncertainty because of the lower number density of source galaxies. 

\subsection{SMACSJ0723 - witnessing the collision of two clusters?}
There are five published strong-lensing maps of SMACSJ0723 \citep{2022mahler, 2022caminha, 2022pascale, 2022golubchik, 2023Sharon}. The five strong-lensing maps are, in general, consistent with each other and show that SMACSJ0732 is elongated in the east-west direction. However, the strong-lensing results derived by \cite{2022mahler} and \cite{2022pascale} are two-component dark matter models with the main halo centered on the BCG. The secondary halo in \cite{2022mahler} is placed in the intracluster light (\mytilde 250~kpc west of the BCG) and the secondary halo in \cite{2022pascale} is placed on the galaxy \mytilde50 kpc to the west of the BCG. The intracluster light, which in theory should be tracing the underlying dark matter distribution, also elongates in the east-west direction with numerous streams and loops \citep{2022montes}. \cite{2022montes} infer from the plumes of the intracluster light that a merger is occurring between the BCG and the galaxy \mytilde 50 kpc to the west.  The weak-lensing mass map determined in our study shows that the dark matter distribution primarily elongates in the east-west direction with a slight extension toward the northeast (see Figure \ref{fig:mass_map}). 


\begin{figure*}
    \centering
    \includegraphics[width=\textwidth]{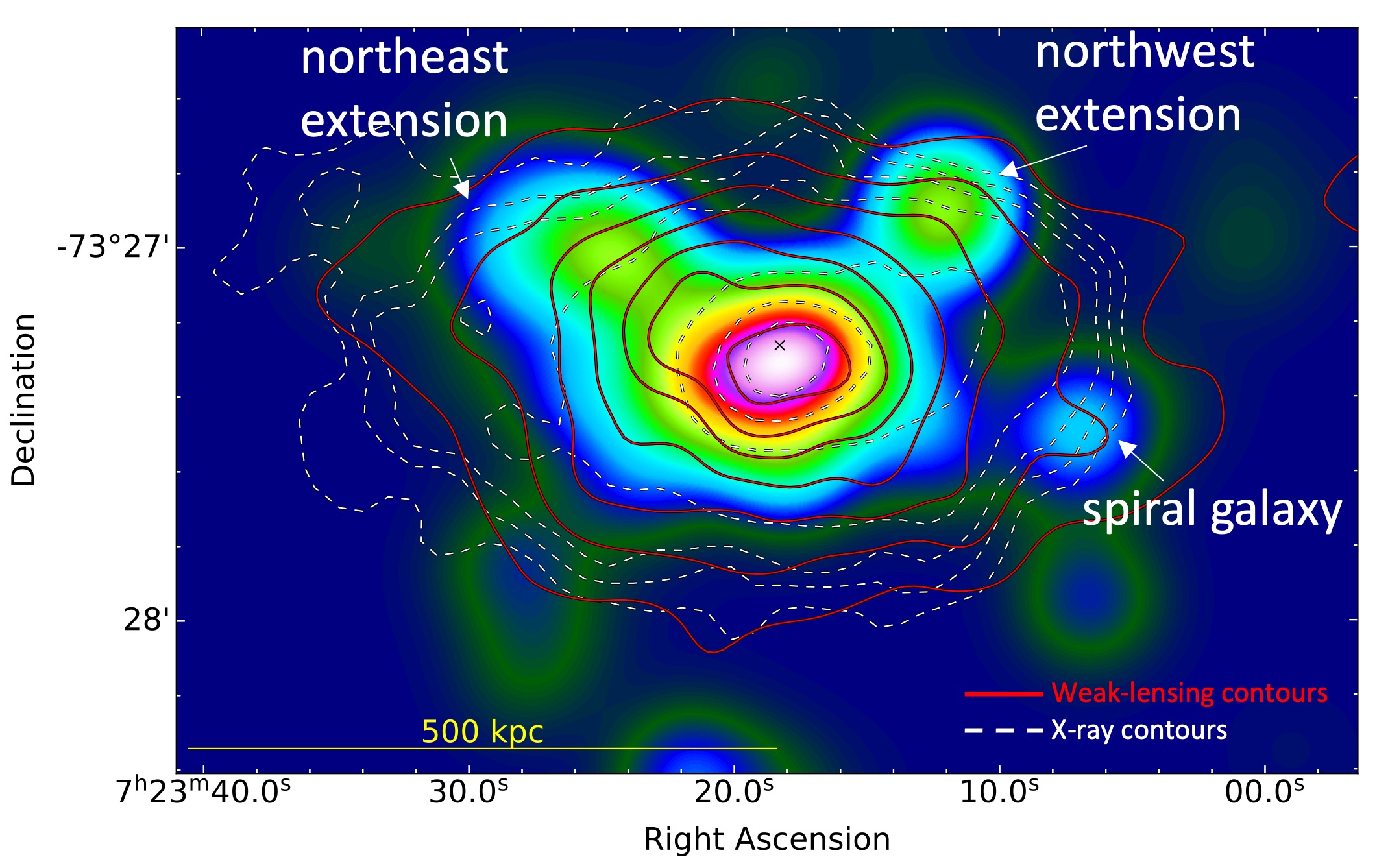}
    \caption{ Weak-lensing mass map (red contours) and X-ray emission (white-dashed contours) overlaid on smoothed ($\sigma=8\arcsec$ Gaussian) luminosity of cluster galaxies. Cluster galaxies includes both spectroscopic members and photometric candidates selected in Section \ref{sec:source_selection} and marked in Figure \ref{fig:cmd}. The northeast extension is visible in the light from the galaxies. The cluster spiral galaxy in the west can also be associated to a bump in the mass distribution. }
    \label{fig:mass_gal}
\end{figure*}

\cite{2022liu} present the eROSITA and \textit{Chandra} X-ray emission of SMACSJ0723. Their X-ray emission maps corroborate that SMACSJ0723 is elongated in an east-west direction. We show the 0.5-7~keV \textit{Chandra} X-ray emission (white-dashed contours) in Figure \ref{fig:mass_gal} with our weak-lensing mass map (red contours). The peak of the X-ray emission (black $\times$) is consistent with the weak-lensing peak. Approximately 450 kpc to the northeast of the X-ray brightness peak is a candidate X-ray substructure that we have labeled the northeast extension. This substructure sits on the boundary of the JWST imaging. However, the extension in the northeast of the mass map is pointed toward the substructure. Wide-field imaging may be able to detect this substructure with weak-lensing.

We show the agreement between galaxies and dark matter in Figure \ref{fig:mass_gal}. The spec-z and photo-z cluster galaxies that were selected in Section \ref{sec:source_selection} are used in creating the luminosity map (colored background map). The peak of the mass map and galaxy luminosity are consistent since the BCG is dominating the total luminosity. The galaxy luminosity extends to the northeast as we noticed in the mass map. We also find a northwest extension in the galaxy light, which, however, is not well traced by the mass map. To the west, the mass contours surround a spiral galaxy with a photo-z estimate that places it in the cluster.  

In Appendix \ref{app:A}, we present a MAXENT mass map generated from our weak-lensing analysis with the archival HST F814W observations ($202\arcsec$ x $202\arcsec$ field of view). The HST mass map also shows the east-west elongation with a slight extension to the northeast but it does not provide substantial evidence for a substructure. The HST mass map has a peak significance of $5.7\sigma$, about half of the JWST value of $12.2\sigma$. By fitting an elliptical Gaussian function to the HST and JWST mass maps, we find that their position angles are consistent, with values that only differ by 4 degrees. 

\textit{Is SMACSJ0723 a merger?}
Combining the information from the strong-lensing measurements of the cluster core, the X-ray and radio emission from the ICM, and our weak-lensing mass maps, we revisit the current state of the cluster. The giant radio halo reported in \cite{2016martinez} and the elongated mass and ICM distributions are indicators of recent merger activity. Additional evidence for a merging system comes from the two-halo models determined from strong lensing \citep{2022mahler, 2022pascale} and the features of the intracluster light. On the other hand, the agreement between the mass peak, galaxy luminosity peak, and X-ray emission peak are evidence toward a more relaxed state. In the following, we theorize that we observe SMACSJ0723 likely in a very early stage of merging. 

\begin{figure*}
    \centering
    \includegraphics[width=\textwidth]{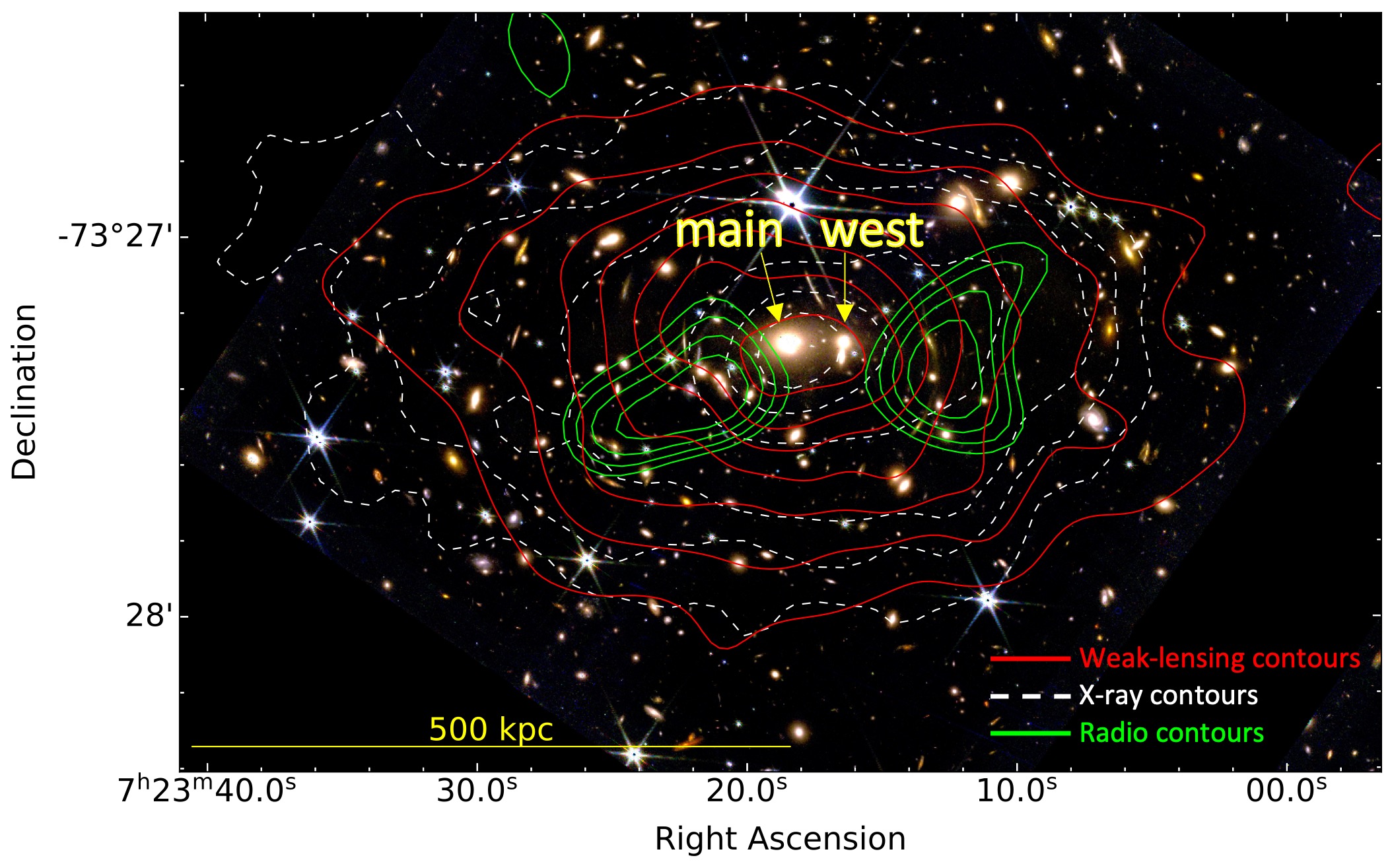}
    \caption{JWST color image with 20ks \textit{Chandra} X-ray (white-dashed contours), weak-lensing mass map (red contours), and RACS 888 MHz radio emission (green contours). The theorized cluster merger is between the \textit{main} and \textit{west} components. The RACS radio emission shows a two distributions that are projected on either side of the two merging components. Their positions are consistent with cluster merger shocks.}
    \label{fig:merger_scenario}
\end{figure*}

The strong-lensing model of \cite{2022pascale} and the intracluster light analysis by \cite{2022montes} both suggest a merger between the main component that is centered on the BCG and the secondary component that is located at the bright galaxy immediately to the west of the BCG (marked in Figure \ref{fig:merger_scenario} as \textit{main} and \textit{west}, respectively). The projected separation (\mytilde 50 kpc) of the \textit{main} and \textit{west} components complicates the merger scenario because of the degeneracy in viewing angle (all angles from plane of sky to line of sight are feasible). The limited number of spec-zs do not show a two-component distribution nor is there a large difference in the redshifts between the BCG and the galaxy at the \textit{west} component. However, \cite{2022mahler} show that the BCG ($z=0.3912$) is at a higher redshift than the systematic redshift of the cluster ($z=0.3877$). Thus, it is difficult to discern whether a merger between \textit{main} and \textit{west} is close to pericenter and merging close to the plane of the sky or has a larger true separation and is merging closer to the line of sight. 

\textit{If SMACSJ0723 is in a merger, what is its current state?} The radio emission may be the best constraint on the state of the system. \cite{2016martinez} reported that the giant radio halo is under-luminous for a cluster of SMACSJ0723 mass. \cite{2013donnert} simulated a cluster merger (total mass $M_{200}=1.5\times10^{15}$ M$_\odot$, similar to the mass of SMACSJ0723) and tracked the evolution of the radio power of the resulting radio halo. From their model, an under-luminous radio halo can be found in the early merging stages (pre-merger to slightly after collision) or in a short period that begins \mytilde1 Gyr post collision and lasts until the halo cools below the radio detection limit (about 0.5 Gyrs later). In the time period between, the radio halo is not expected to be under-luminous. If we naively place the 1.4~GHz radio power ($1.72\pm0.22\times10^{31}$ erg s$^{-1}$ Hz$^{-1}$) from \cite{2016martinez} and the X-ray luminosity ($11.2^{+0.2}_{-0.1}\times10^{44}$ erg s$^{-1}$) from \cite{2022liu} onto the radio power as a function of X-ray luminosity relation from figure 4 of \cite{2013donnert}, we expect a cluster merger that is near first pericenter passage. However, \cite{2013donnert} also show that the pre- and post-merger states can be distinguished by the radio spectral index with a flat spectrum occurring during the early stages of the merger and a steeper spectrum found in later stages due to synchrotron aging. Thus, a spectral index would be a more useful method to identify the phase of the merger. With the ATCA observation, \cite{2016martinez} was unable to measure the spectral index. Deep \textit{Chandra} observations could also test the first pericenter hypothesis by searching for the temperature increase caused by compression of the ICM during the collision. 

An alternative is that the radio emission is not a giant radio halo, but instead is a pair of radio relics \citep[see][for a review]{2019vanweeren}, which are synchrotron emission from merger-induced shocks. We retrieved the 888 MHz Rapid ASKAP Continuum Survey (RACS)-low1 DR1 data \citep{2021racs} from the CASDA. The data is convolved to a $25\arcsec$ resolution. Figure \ref{fig:merger_scenario} shows the radio emission contours (green), weak-lensing mass contours (red), and \textit{Chandra} contours (white dashed) overlaid on the JWST color image. The RACS data displays two diffuse radio distributions that are found on opposing sides of the two bright cluster galaxies (\textit{main} and \textit{west}). This is the configuration that is expected for a recent collision of two clusters. We propose that the radio emission is candidate radio relics. However, as \cite{2018ha} have shown, the kinetic energy flux shortly after collision is low and bright radio relics are not expected until about \mytilde0.8 Gyr after collision. Therefore, if the diffuse radio emission is from radio relics, a significant line of sight component to the merger would be expected. Higher resolution radio observations would be able to define the morphology of the radio emission, which would aid in determining its nature. Also, a spectral index map would be useful for classifying the radio emission. If the diffuse radio emission is double radio relics, their projected separation (\mytilde200 kpc) would be the shortest observed to date and the SMACSJ0723 system could be an important glimpse into an early stage of cluster mergers.

\section{Conclusions} \label{sec:conclusions}
We have conducted the first weak-lensing analysis with the James Webb Space Telescope. Our weak-lensing analysis investigates the massive galaxy cluster SMACS J0723.3–7327, which was one of the Early Release Observations. We performed our analysis in the F200W filter.  

Using stars as a probe of the PSF, we investigated the ellipticity of the PSF and its spatial variation in the SMACSJ0723 imaging. Our measurements found that for both modules the PSF is predominantly stretched in the x-direction of the detectors with very small spatial and temporal variation. On average the PSF has measured ellipticity components of $e_1=0.007\pm0.001$ and $e_2=0.029\pm0.001$.
~We performed a principal component analysis of the stars and used the result to model the spatial dependence of the PSF. Our PSF models properly account for the ellipticity of the PSF and leave minimal residual shapes. By comparing the shapes of our PSF models to that of the stars, we found mean residual ellipticities of $e_1=(0.3\pm3.5)\times10^{-4}$ and $e_2=(1.8\pm4.0)\times10^{-4}$. The mean residual size after PSF subtraction is $(7.0\pm11.7)\times10^{-2}$ mas.    

We measured the shapes of galaxies by forward modeling the PSF and fitting the elliptical Gaussian function. Utilizing the publicly available redshift catalog, we selected background galaxies by their photometric redshift and compiled them into a source catalog for weak lensing. The source catalog has \mytilde 215 galaxies arcmin$^{-2}$, approximately 5 times more than HST for the same field.

From the source catalog, we mapped the mass distribution of SMACSJ0723 using a maximum entropy technique. We detect the galaxy cluster peak at the $12.2\sigma$ level. The convergence map highlights the east-west elongation that other studies have found. It also extends to the northeast toward a candidate substructure that is present in the \textit{Chandra} X-ray image. We derive the mass centroid uncertainty through bootstrap realizations of the mass map and find the mass centroid to be consistent with the BCG and the X-ray brightness peak at the $1\sigma$ level.

To estimate the mass of the cluster, we calculated the tangential shear profile centered on the BCG. This analysis shows the advantage of having two modules on the JWST because the second module probes the cluster outskirts near $R_{500}$. By fitting an NFW model to the tangential shear, we found the mass of the cluster to be $M_{500}=7.9\pm1.1\times10^{14}$ M$_{\odot}$ ($M_{200}=11.4\pm1.5\times10^{14}$ M$_\odot$ ). This mass estimate is consistent with the X-ray and SZ mass estimates from literature. 

To understand whether MACSJ0723 is undergoing a merger, we combine the multiwavelength analyses from past studies with our weak-lensing result. The cluster has multiple features that signify that it is merging. The weak-lensing mass map, the strong-lensing mass maps, the X-ray emission, and the galaxy distribution are elongated. Furthermore, the cluster hosts diffuse radio emission that has been classified as a giant radio halo. Arguments against an ongoing merging come from the good spatial agreement between the gravitational lensing peaks, the X-ray brightness peak, and the galaxy luminosity peak. By combining the multiwavelength probes, we show that the observed configuration is consistent with a merger that is approaching or just past first pericenter passage. We also show that, in the RACS 888 MHz radio band, the diffuse radio emission is found to have two components that are equidistant from the core of the cluster and we suggest that they could be radio relics instead of a radio halo. Additional radio observations would be immensely helpful to constrain the stage of the merger and the nature of the diffuse radio emission.

We have shown the remarkable power of JWST as a weak-lensing telescope to study the properties of galaxy clusters. The detector resolution and filters lead to a high number density of source galaxies for weak-lensing, which provides tight statistical constraints on the mass and high weak-lensing signal detection significance. The two modules are useful for simultaneously probing the cores of galaxy clusters and constraining the mass profile at larger radii. 


\begin{acknowledgments}
K.F. would like to thank Bomee Lee for discussion on photometric redshift estimation.
M. J. Jee acknowledges support for the current research from the National Research Foundation (NRF) of Korea under the program 2022R1A2C1003130.

\end{acknowledgments}

%

\vspace{5mm}
\facilities{JWST(NIRCam), \textit{Chandra}(ACIS), HST(ACS), ASKAP}


\software{astropy \citep{astropy:2013, astropy:2018, astropy:2022},  
          Source Extractor \citep{1996bertin}, Colossus \citep{2018diemer},
          scipy \citep{2020scipy}, Grizli \citep{2019grizli}}

\appendix
\section{HST ACS F814W weak lensing} \label{app:A}
We provide an HST view of the weak-lensing signal of SMACSJ0723. The archival F814W imaging (2529~s of integration, PI: Ebeling) was downloaded and processed to a coadded mosaic with our data reduction pipeline \citep[pipeline steps are outlined in][]{2017finner, 2021finner, 2023finnera}. A catalog of objects was created with SExtractor. Utilizing the \cite{2007bjee} publicly available PSF templates, we generated PSF models for each of the galaxies in the catalog. Following the same procedure as taken for the JWST analysis, we selected background galaxies based on their photometric redshift and shape measurements (Section \ref{sec:source_selection}). The resulting source catalog contains 45 galaxies arcmin${^{-2}}$ over the ACS field of view. For the source catalog, the mean $\beta$ and $\beta^2$ are 0.53 and 0.33, respectively.  

\begin{figure*}
    \centering
    \includegraphics[width=\textwidth]{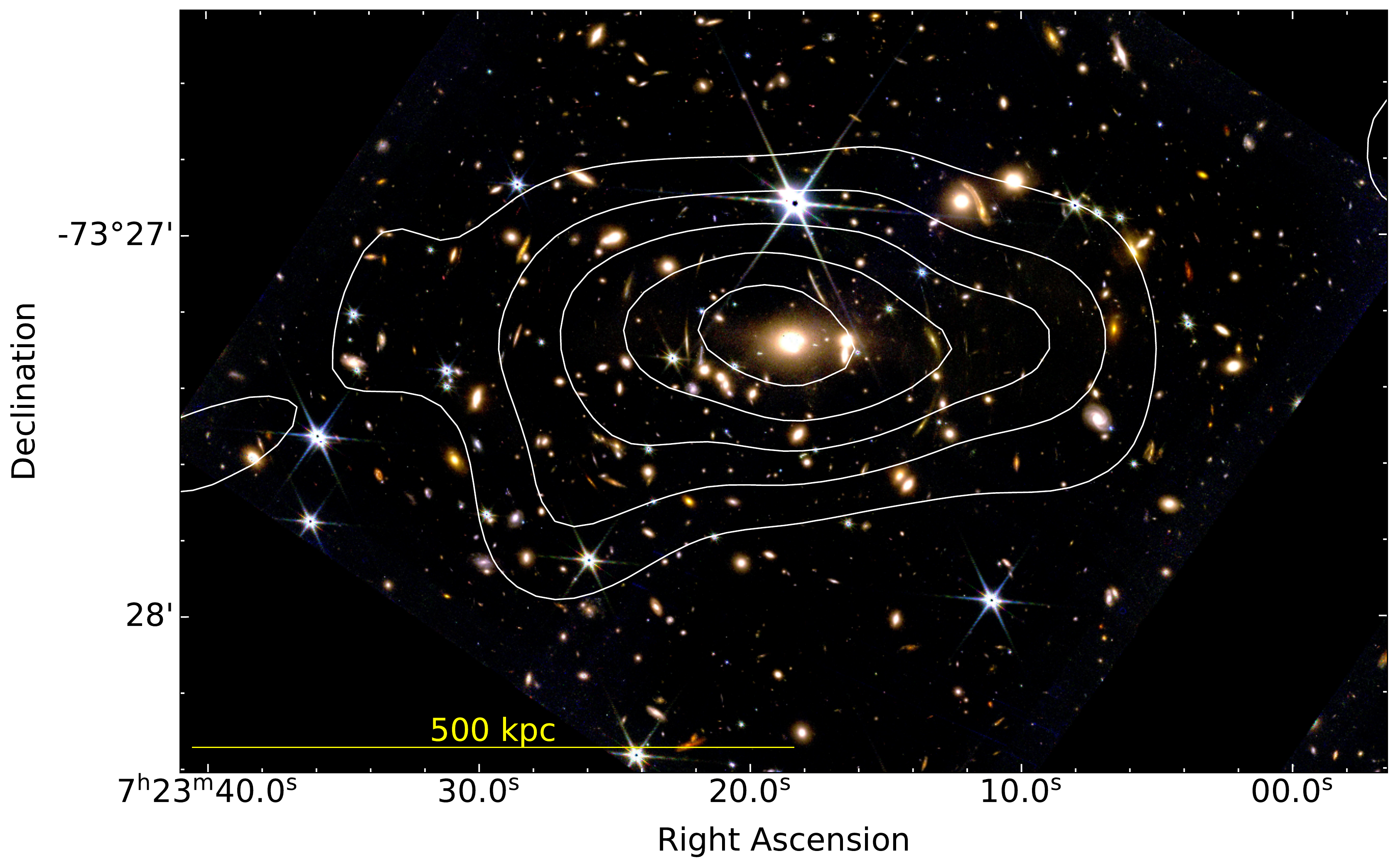}
    \caption{Projected mass distribution (white contours) derived from the HST imaging of SMACSJ0723. The contours start at $1\sigma$ and increase at intervals of $1\sigma$. As seen in the JWST version and the strong-lensing maps, the mass distribution is elongated in the east-west direction.}
    \label{fig:hst_mass_map}
\end{figure*}

Figure \ref{fig:hst_mass_map} presents the projected mass distribution derived from HST imaging over the JWST color image. The projected mass distribution is detected at the $5.7\sigma$ level (compared to $12.2\sigma$ from JWST). In general, the mass distribution is similar to the JWST version with an east-west elongation.

\begin{figure}
    \centering
    \includegraphics[width=0.5\textwidth]{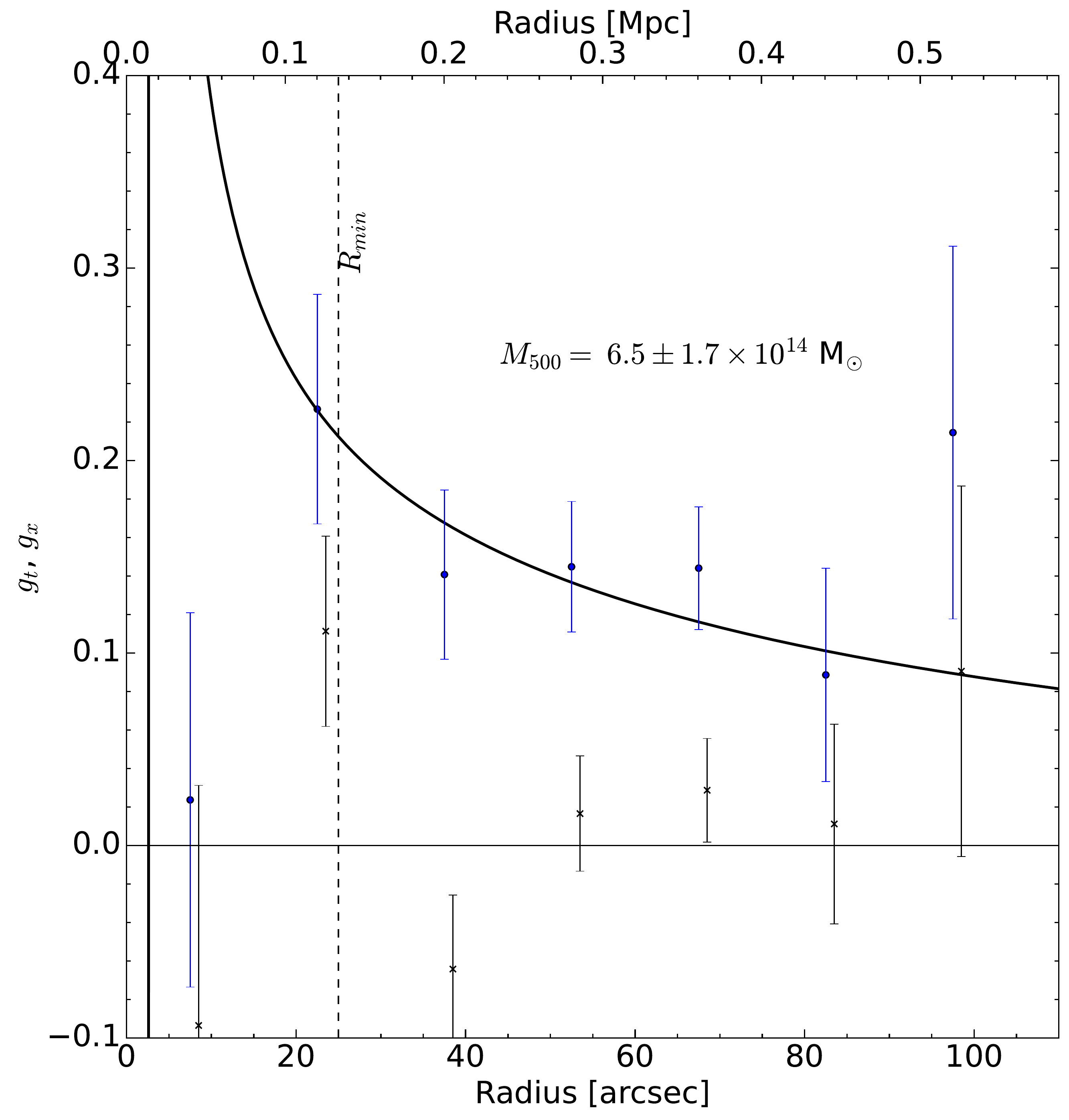}
    \caption{Tangential (blue circles) and cross (black x's) shear measurements from the HST F814W imaging. Galaxies inside $R_{min}$ are excluded from the fit because they are within the strong-lensing regime. The NFW fit (black line) estimates the mass of the cluster to be $M_{500}=6.5\pm1.7\times10^{14}$ M$_\odot$.}
    \label{fig:hst_tan_shear}
\end{figure}

Figure \ref{fig:hst_tan_shear} shows the tangential shear measured from the cluster center (BCG). As we did for the JWST version, we excise the inner $25\arcsec$ to remove the strong-lensing region. The NFW fit to the HST tangential shear provides a mass estimate of $M_{500}=6.5\pm1.7\times10^{14}$ M$_\odot$ ($M_{200}=9.4\pm2.4\times10^{14}$ M$_\odot$), which is consistent with the JWST measurement.

\bibliography{sample631}{}
\bibliographystyle{aasjournal}



\end{document}